\documentclass[12pt]{article}
\usepackage[affil-it]{authblk}
\usepackage{etex}
\usepackage[utf8]{inputenc}
\usepackage[T1]{fontenc}
\usepackage{textcomp}
\usepackage[a4paper,tmargin=3cm,bmargin=3cm,rmargin=3cm,lmargin=3cm]{geometry}
\usepackage{lmodern}
\usepackage[normalem]{ulem}
\usepackage{enumitem}

\usepackage{amsfonts,amssymb,amsmath,amscd,mathtools,cite,slashed}
\usepackage[amsthm,thmmarks,amsmath]{ntheorem}
\usepackage{graphicx}
\usepackage[all]{xy}
\usepackage{lmodern}
\usepackage{tikz}
\usetikzlibrary{shapes.misc,arrows,decorations.markings}

\usepackage{color}
\usepackage{psfrag}
\usepackage{epsfig}
\usepackage{pdflscape}
\usepackage{hyperref}
\usepackage{verbatim} 
\usepackage{graphicx}

\usepackage{xcolor}
\usepackage{framed}
\definecolor{shadecolor}{gray}{0.8}
\definecolor{lgray}{gray}{0.5}

\newcounter{mnotecount}[section]
\renewcommand{\themnotecount}{\thesection.\arabic{mnotecount}}
\newcommand{\mnote}[1]
{\protect{\stepcounter{mnotecount}}$^{\mbox{\footnotesize
$
\bullet$\themnotecount}}$ \marginpar{
\raggedright\tiny\em
$\!\!\!\!\!\!\,\bullet$\themnotecount: #1} }

\definecolor{darkgreen}{rgb}{0,.5,0}

\theoremstyle{definition}

\newcommand{\norm}[1]{\left\Vert {#1} \right\Vert}

\newcommand{\sR}{{\mathbb R}}

\newcommand{\M}{\mathcal{M}}

\renewcommand{\P}{\mathcal{P}}

\newcommand{\T}{\mathcal{T}}

\newcommand{\R}{\mathcal{R}}

\DeclareMathOperator{\supp}{supp}

\newcommand{\vc}{\vcentcolon =}

\renewcommand{\R}{\mathbb{R}}

\newcommand{\ev}{\mathrm{ev}}

\begin{document}

\title{Causality and time order -- relativistic and probabilistic aspects}
\author{Micha\l\  Eckstein ${}^{1,2}$}
\author{Michael Heller ${}^{2,3}$}

\affil{\small ${}^1$ Institute of Theoretical Physics, Jagiellonian University, ul. prof. Stanis{\l}awa {\L}ojasiewicza 11, 30-348 Krak\'ow, Poland}

\affil{\small ${}^2$ Copernicus Center for Interdisciplinary Studies, ul. Szczepa\'nska 1, 31-016 Krak\'ow, Poland}

\affil{\small ${}^3$ Vatican Observatory, V-00120 Vatican City State}

\maketitle

\begin{abstract}
We investigate temporal and causal threads in the fabric of contemporary physical theories with an emphasis on empirical and operationalistic aspects. Building on the axiomatization of general relativity proposed by J. Ehlers, F. Pirani and A. Schild (improved by N. Woodhouse) and the global space-time structure elaborated by R. Penrose, S.W. Hawking, B. Carter and others, we argue that the current way of doing relativistic physics presupposes treating time and causality as primitive concepts, neither of them being `more primitive' than the other. The decision regarding which concepts to assume as primitive and which statements to regard as axioms depends on the choice of the angle at which we contemplate the whole. This standard approach is based on the presupposition that the concept of a point-like particle 
 is a viable approximation. However, this assumption is not supported by a realistic approach to doing physics and, in particular, by quantum theory. We remove this assumption by analysing the recent works by M. Eckstein and T. Miller. They consider the space $\P(M)$ of probability measures on space-time $M$ such that, for an element $\mu \in \P(M)$, the number $\mu (K)$ specifies the probability of the occurrence of some event associated with the space-time region $K$ and the measure $\mu$. In this way, $M$ is not to be regarded as a collection of space-time events, but rather as a support for corresponding probability measures. As shown by Eckstein and Miller, the space $\P(M)$ inherits the causal order from the underlying space-time and facilitates a rigorous notion of a `causal evolution of probability measures'. We look at the deductive chains creating temporal and causal structures analysed in these works, in order to highlight their operational (or quasi-operational) aspect. This is impossible without taking into account the relative frequencies and correlations observed in relevant experiments.
\end{abstract}

\section{Introduction}
It is perhaps commonplace to attribute the relational concept of time to Leibniz. Every philosopher knows that, according to Leibniz, time is but a relation ordering events one after the other. It is less known that at the end of his life Leibniz supplemented his relational conception of time with what later resulted in the causal theory of time\footnote{The note concerning this conception is found in Leibniz's essay entitled `The metaphysical foundations of mathematics' \cite{Leibniz1} and published posthumously. The present section is based mainly on chapter one of the extensive study by Henry Mehlberg devoted to the causal theory of time \cite{Mehlberg}.}. His new idea attempted to identify the nature of relations constituting the time order. When we read Leibniz on ordering relations, we should not ascribe to him our present concept of formal order, since his own aim was to stress the difference between his own understanding and that of his English opponents (Newton and Clarke). In the `English theory', there exist two classes of entities: instances and events, and the `natural order' of instances\footnote{We would today say the order determined by the metric structure of time.} determines the order of events. Events happen at certain instances, but instances are independent of events.  In Leibniz's approach, the events are the only class of entities, and time is a derived concept given by relations ordering events. The causal conception of time adds a new idea to Leibniz's philosophy. In his metaphysical-literary style: `the present is always pregnant with the future' \cite[p. 557]{Leibniz2}, and as explained by Mehlberg: `\ldots if one arranges phenomena in a series such that every term contains the reason for all those which come after it in the series, the causal order of the phenomena so defined will coincide with their temporal order of succession.' \cite[p.~46]{Mehlberg}. Leibniz's idea contrasts with that of Hume who attempted to reduce causal order to temporal order. Whitrow puts this in the following way: in Hume's view `the only possible test of cause and effect is their `constant union', the invariable succession of the one after the other' \cite[p. 323]{Whitrow}.

The advent of the theory of relativity showed the role played by time and causality in the structure of modern physics in a radically new light and has had a profound impact on our understanding of the world. The aim of the present study is to confront the traditional dispute concerning the mutual relationship between time and causality with what contemporary physical theories import to this issue with an emphasis on  the empirical and operationalistic aspects. Our approach consists not so much in analysing separate problems involved in time--causality interaction, but rather in investigating temporal and causal threads in the fabric of contemporary physical theories. 

The structural construction of a physical theory is best visible in its axiomatisation. We thus start with an axiomatization of the theory of relativity. Perhaps the earliest such axiomatization is attributed to A. A. Robb \cite{Robb} who in 1914 proposed an axiomatic system based on the concept of `conic order', and was able to derive both topological and metric properties of space-time from the `invariant succession of events'. In the axiomatic systems of both Carnap \cite{Carnap} and Reichenbach \cite{Reich} it is the temporal order that is reduced to that of causal order. The same is true for the Mehlberg's  system \cite{Mehlberg} which was elaborated within the broader setting of an interdisciplinary study of the causal theory of time\footnote{Although Mehlberg's book appeared in 1980, it is based on his works dating back to 1935 and 1937.}.

The latter three authors were associated with logical empiricism and, in agreement with its philosophical ideology, aimed at clarifying logically the conceptual situation in the theory of relativity, one which was extensively discussed at that time. In doing so, they eliminated, with the help of their axioms, some `pathological situations' (such as the existence of closed timelike curves). What they did not take into account was that the later development of physics might need such pathologies, if not for empirical reasons then as auxiliary hypotheses for proving some general theorems. This became evident when in 1949 Kurt G\"odel \cite{Godel} published his solution to Einstein's field equations with closed timelike curves (soon after more solutions with similar `time anomalies' were found). A few years later the first general theorem concerning the global structure of space-time was proved stating that every compact space-time must contain closed timelike curves \cite{BassWitten}.

In the present study, we take into account both these lines of investigation. We adopt the axiomatization proposed by J. Ehlers, F. Pirani and A. Schild \cite{Ehlers}  which is considered the most adequate axiomatization of general relativity (in the following, we refer to it as to EPS axiomatization). Its purpose is not so much to `clarify inter-conceptual relations', as it was in the case with previous axiomatizations, but rather to reveal the structuring of space-time (i.e. showing mutual relations between its various substructures). In Section \ref{sec:arch}, we briefly present the EPS system with an emphasis on its quasi-operational aspect that we then  exploit when discussing the empirical anchoring of space-time physics.

The second thread of investigation, mentioned above, resulted into the so-called global study of space-time structures. Within this approach the causal structure of space-time has been worked out in great detail (see, for instance, \cite{Carter71}, \cite[Chapter 6]{HawkingEllis}). This approach entered into relativistic physics as a part of its theoretical tool-kit. These tools were used by N. M. J. Woodhouse \cite{Wood} to improve the EPS axiomatization. In Section \ref{sec:global}, we refer to some aspects of this approach, mainly to prepare the conceptual environment for our further analysis.

In Section \ref{sec:primitve}, we draw some conclusions from the above analysis. The current way of doing relativistic physics presupposes treating time and causality as `primitive concepts'. They are mutually interrelated, but neither of them is `more primitive' than the other, and the decision which concepts to assume as primitive and which statements to regard as axioms depends on the choice of the angle at which we contemplate the whole.

Everything said above is based on the presupposition that the concept of a point-like space-time event is a viable approximation. However, such an assumption cannot be fulfilled in actual physical experiments. Moreover, in a fundamentally probabilistic theory, such as quantum mechanics, there are no definite events at all. Even in classical theory space and time localization measurements are affected by experimental uncertainties, and in quantum theory statistical predictions apply also to single events. In Section \ref{sec:prob}, we take this into account by pondering the space $\mathcal{P}(M)$ of all probability measures on space-time $M$. For any element $\mu \in \mathcal{P}(M)$ and a space-time region $K$ the number $\mu (K)$ is the probability of the occurrence of some event. 
In this way, $M$ is not to be regarded as a collection of space-time events, but rather as a support for corresponding probability measures.

The question arises: when do `probabilistic uncertainties' propagate causally in space-time? More precisely, when can the detection statistics $\mu $ causally influence the detection statistics $\nu $? To answer this question we briefly present, in Section \ref{sec:prob_caus}, a generalization of the relativistic causal structure to `events' spread in space-time and represented by suitable probability measures. This generalisation was recently proposed by Eckstein and Miller \cite{AHP2017}. For two probability measures $\mu, \nu \in \P(M)$, the measure  $\mu$ is said to causally precede $\nu$, symbolically $\mu \preceq \nu$, if for every compact set $K \subset \supp \mu$ we have $\mu(K) \leq \nu (J^+(K))$, where $J^+(K)$ is the causal future of $K$.

In order to discuss the time-evolution of such measures meaningfully one has to specify a time parameter. Because the measures are inherently non-local objects, one needs to chose a global time-function $\T: M \to \R $. To this end, one restricts oneself to the class of globally hyperbolic spacetimes which admit well-posed evolution problems with initial data specified on a Cauchy hypersurface $S$. Then, one considers   a family of probability measures supported on the corresponding time-slices, $\mu_t \in \P(S_t)$, with $S_t = \{t\} \times S \subset M$ and $t \in \R$. 
Quite naturally, one says that a time-evolution of measures $\{\mu_t\}_t$ is causal if $\mu_s \preceq \mu_t$ for all $s \leq t$. Remarkably, as shown by Miller \cite{Miller17a,Miller21}, there exists an invariant object, call it $[\sigma ]$, which is defined in terms of probability measures on worldlines in a given hyperbolic space-time, and the evolution of measures $\{\mu_t\}_t$ is causal if and only if such an invariant object exists. Some philosophical aspects of this formalism are discussed in Section~\ref{sec:prob_evo}.

Finally, in Section \ref{sec:Conclusions}, we collect some comments regarding the analyses carried out in this work. Our guiding idea was to preform all these analyses from an operational or, where this was not possible, `operational-in-principle' point of view. In the final section we look once again  at the deductive chains of our reasoning and the conclusions to which they lead, in order to highlight the operational, or quasi-operational, aspect of them. As we have tried to show, certain fundamental properties of time, space and causality and their interrelationships are enforced by the very method of physics and by the fact that its very essence consists in its empirical approach to the reality under study. It is in the interest of this method that the quasi-operational elements of the method be replaced, as far as possible, by truly operational elements. The basic methodology of experimental physics is based on  probabilistic concepts. Therefore, the generalization of the relativistic causal structure to probabilistic measures, presented in the final sections of this work, is to be welcomed as a step in the right direction.

\section{Space-Time Architecture\label{sec:arch}}

From the mathematical point of view, a space-time in general relativity is a pair $(M,g)$, where $M$ is a four-dimensional differential manifold, and $g$ a Lorentzian metric defined on it. Furthermore, it is typically assumed that $(M,g)$ is connected and time-oriented \cite{Wald}.

We say that the manifold $M$ carries a Lorentzian structure. This structure not only contains several other mathematical substructures which interact with each other, creating a subtle hierarchical edifice, but also admits, on each of its levels, a physical interpretation, making out of the whole one of the most beautiful models of contemporary  physics. We shall briefly describe the `internal design' of this model. Our analysis will be based on the classic paper by Ehlers, Pirani and Schild \cite{Ehlers} in which these authors presented a quasi-operationistic axiomatic system for the Lorentz structure of space-time showing both its mathematical architecture and physical meaning.

The building blocks (primitive concepts) of this axiomatization are: (1) a set $M = \{p, q, \ldots \}$, the elements of which are called events; and two collections of subsets of $M$: (2) $L = \{L_1, L_2, \ldots \}$, the elements of which are called histories of light rays or of photons (light rays or photons, for the sake of brevity); (3) $P = \{P_1, P_2, \ldots \}$, the elements of which are called histories of test particles or of observers (particles or observers, for brevity).

The axioms guarantee the following situations. A light is sent from an event $p$, situated on a particle history $P_1$, and received  at an event $q$, situated on a particle history $P_2$ (a message from $p$ to $q$). The message can be reflected at $q$ and sent to the event $p'$, situated on the particle history $P_1$ (an echo on $P_1$ from $P_2$). By a suitable combination of messages and echoes one can ascribe four coordinates to any event, and construct local coordinate systems. Some mathematical gymnastics, sanctioned by suitable axioms, allows one to organise the set of events, $M$, into a differential manifold (with the usual manifold topology).

The next set of axioms equips the manifold $M$ with a conformal metric. This is the usual metric (with a Lorentz signature) defined up to a factor of proportionality (conformal factor). This metric allows one to distinguish timelike, null (lightlike) and spacelike histories (curves), and completely determines the geometry of null-curves.

To determine the geometry of timelike curves one needs additional axioms defining the so-called projective structure. Its task is to determine a distinguished class of timelike curves that physically are interpreted as representing histories of particles (or observers) moving with no acceleration (freely falling particles or observers).

Conformal and projective structures are, in principle, independent and they need to be synchronised. This is done with the help of suitable axioms which enforce null geodesics passing through an event $p$ to form the light cone at $p$ and projective timelike curves to fill in the interior of this light cone. If this is the case, we speak of the Weyl structure.

In a Weyl space (a manifold equipped with the Weyl structure), there is a natural method to define length -- the arc length -- along any timelike curve. Such a length is interpreted as a time interval measured by a clock carried  by the corresponding particle (proper time of the particle). However, proper times of different particles are unrelated; to synchronise them a suitable metric must be introduced on $M$. This can be done with the help of the following axiom. Let $p_1, p_2,...$ be equidistant events on the history of a freely falling particle $P$, and let they be correspondingly simultaneous with events $q_1, q_2,...$ on the history of a freely falling particle $Q$. Simultaneity is understood here in the Einstein sense: two events, $p$ and $q$, are simultaneous if an observer situated half-way between them sees the light signal emitted by $p$ and $q$ at the same instant as shown by the observer's clock. The metric structure is established if the events $q_1, q_2,...$ on $Q$ are also equidistant. The metric $g$, constructed in this way, contains in itself the Weyl structure; it is, therefore a Lorentz metric. This completes the construction of the relativistic model of space-time.

\begin{figure}[h]
\begin{center}
\includegraphics[scale=1.35]{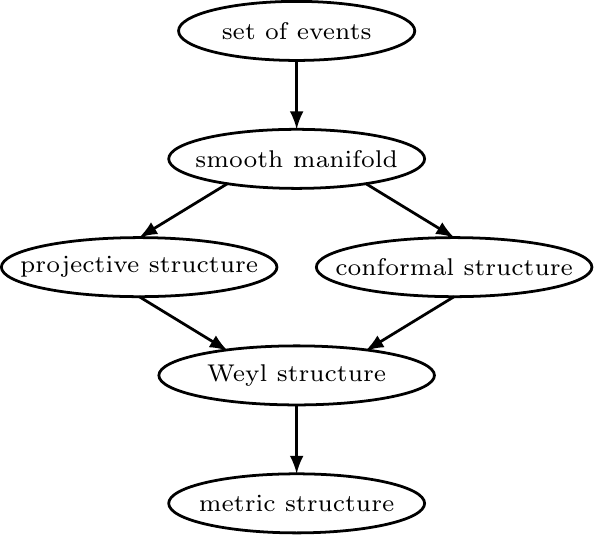}
\caption{\label{Fig1}A diagram illustrating the subsequent layers of axiomatisation of general relativity following \cite{Ehlers}.}
\end{center}
\end{figure}

It is important to stress that the axioms of general relativity, as formalised in \cite{Ehlers}, can and have been tested against the empirical data. For example, the axiom on the conformal structure of space-time would be violated if the velocity of photons would depend upon its energy \cite{DSR,QGPhenomenology}. No such effect has been observed to a high degree of accuracy \cite{DSRFermi,QG_astro_constraints_2015,GRB20}. In the same vein, one could seek deviations from the universality of the projective structure by inspecting the dispersion relations of high-energy massive particles \cite{DSR_neutrinos}.

The existence of the Lorentzian structure on top of the Weyl structure could be undermined through the observation of the `second clock effect' \cite{Ehlers}. The latter arises commonly in various modified-gravity theories (see e.g. \cite{fR_LRR}). Recenly, the second clock effect was constrained using the CERN data on the muon anomalous magnetic moment \cite{2clockCERN}.

Eventually, one can question the model of a space-time as a smooth manifold. This is typically done on the basis of some quantum gravity theory \cite{QGreview}. Although the possible effects of these models are extremely hard to observe, the experimental efforts in this direction are progressing \cite{QGPhenomenology}. In particular, some limits on space-time granularity were recently established \cite{Holometer_res2016}.

\section{Causal structure and global time\label{sec:global}}

Every text-book on relativity tells us that causality (the causal structure) is identical with the cone structure of space-time, that is to say with the Weyl structure of the above scheme (synchronised projective and conformal structures). The conformal class of space-time metrics, $[g]$, induces a classification of tangent vectors at any event $p \in M$. A vector $v \in T_p M$ is timelike if $g(v,v) < 0$, spacelike if $g(v,v) > 0$, null  if $g(v,v) = 0$ and causal if $g(v,v) \leq 0$, where $g$ is any representant of the class $[g]$. The set of timelike vectors has two components: the `future' and the `past'. It is standard to assume \cite{Wald} that the space-time is time-oriented (and not only time-orientable), which means that a choice of time-direction is fixed (smoothly) for all vectors throughout the tangent bundle $TM$. A piece-wise smooth curve $\gamma: I \to M$, with an affine parameter $t \in I \subset \R$, is future-directed timelike/causal if $\gamma'(t)$ is future-directed timelike/causal, wherever the vector $\gamma'(t)$ is defined.

The Weyl structure of spacetime induces the chronology and causal relations between the events in $M$. An event $p$ is said to chronologically (causally) precede an event $q$, written $p \ll q$ ($p \preceq q$, respectively), if there is a piece-wise smooth future-directed timelike (causal) curve from $p$ to $q$ \footnote{For the full account of chronology and causal relations see \cite{MS08,MingRev}.}. 
Both of them determine channels through which causal influences can propagate rather than actual interactions between cause and effect. In particular, note that a causal piece-wise smooth curve can be composed of timelike and null segments. In terms of the EPS conceptual framework such a curve would correspond to a signal, which is, e.g., initially encoded in a massive particle and then gets transferred -- via some interaction -- to a photon.

With the help of these relations we define the chronological future $I^+(p)$ and the chronological past $I^-(p)$ of an event $p$ as the set of all events which chronologically follow (resp. are followed by) $p$; and analogously, the causal future $J^+(p)$ and the causal past $J^-(p)$ of $p$. 

If $M$ is a space-time manifold, the tangent space at each of its points (events), $T_pM$ has the structure of the Minkowski space. Its causal structure is the familiar light cone structure of special relativity. This structure, essentially unchanged, is inherited by any local neighbourhood (called a normal neighbourhood) of the space-time manifold $M$.\footnote{Through the so-called exponential mapping.} However, globally (outside normal neighbourhoods) causal structure can be very different from that of Minkowski space, sometimes extremely exotic and full of pathologies (see \cite{Carter71,MingRev}). 

There exists an interaction between the causal structure of space-time and its topology. Sets $I^+(p)$ and $I^-(p)$ are always open, but sets $J^+(p)$ and $J^-(p)$ are not always closed. This allows one to define topology `innate' for causal spaces (i.e. defined entirely in terms of causal relations). It is the so-called Alexandrov topology\footnote{Sets are defined to be open in this topology if they are unions of the sets of the form $I^+(p) \cap I^-(p)$.}. This topology is weaker than the usual manifold topology\footnote{Alexandrov topology coincides with manifold topology if the strong causality condition is satisfied; see below.}.

This mathematical apparatus proved to be very effective in disentangling various problems related to the structure of space-time. Engaging it into subtleties of space-time architecture allowed Woodhouse to improve the EPS axiomatization \cite{Wood}. He was able to derive the differential and causal structures from EPS-like axioms expressed in terms of chronology and causality relations. The main advantage of the Woodhouse approach is that no assumption has to be made concerning paths along which light signals propagate. They are deduced from statements regarding the emission and absorption of light signals. In his approach, one has to assume that there is exactly one history of a particle through each point in each direction in space-time. Physically this means that it is possible to define the history of a freely falling particle through any point in space-time.

The above conceptual machinery provides a powerful tool for studying various global aspects of space-time. In what follows, we focus on those of them that elucidate mutual dependencies between time and causality.

The first condition that has to be implemented in order to have something resembling a temporal order is to guarantee the absence of closed timelike and causal curves. The corresponding conditions are called the chronology condition (the absence of closed timelike curves) and the causality condition (the absence of closed causal curves), respectively. The motivation is obvious: with such loops there is no clear distinction between the future and the past. Although the existence of relativistic world models with closed timelike curves (such as the G\"odel's famous model \cite{Godel}) shows that the idea of `closed time' is not a logical contradiction, yet the point is that the space-time structure strongly interacting with the rest of physics might be a source of many logical perplexities. Not only in a space-time with closed timelike curves one might kill one's ancestor to prevent one's birth, but also --- more prosaically --- in a space-time with causality violations no global Maxwell field could exist that would match a given local field \cite{GerochHorowitz}.

There is a rich hierarchy of stronger and stronger conditions\footnote{In fact, the hierarchy is nondenumerable.} which improve temporal and causal properties \cite{Carter71,HawkingEllis,MS08}. Among them there is an important condition, called the strong causality condition, that excludes almost closed causal curves\footnote{It states that each neighbourhood of any event in space-time contains a neighbourhood which no causal curve intersects more than once.}. In such a space-time, the nonexistence of closed timelike curves is satisfied with a certain safety margin. Owing to this margin, the Alexandrov topology improves to the manifold topology. 

However, this is not enough. The fact that any measurement can only be done within certain unavoidable error limit, prevents us from measuring the space-time metric exactly. Many `nearby metrics' are always within the measurement `box error'. If measurements are to have any meaning at all -- and the very existence of physics depends on this -- we must postulate a certain stability of measurements. This postulate as regarding causality assumes the form of the stable causality condition; it precludes small perturbations of space-time metric to produce closed causal curves\footnote{To define this condition precisely we should consider the space Lor($M$) of all Lorentz metrics on a given space-time manifold $M$, and equip Lor($M$) with a suitable topolgy. Only then  are we able to determine what a small perturbation of a metric means \cite{Hawking1968}.}.

The condition of stable causality therefore has a certain philosophical significance. When it is not satisfied, measurements of physical quantities become meaningless. As Hawking put it, `Thus the only properties of space-time that are physically significant are those that are stable in some appropraite topology' \cite{Hawking1971}. 

It is a nice surprise that this condition of `physical reasonability' meets with another very `reasonable' property. To Hawking we owe the following theorem: In a space-time $M$ there exists global time, measured by a global time function, if and only if $M$ is stably causal \cite{Hawking1968}. The history of any clock in the universe (for instance, the history of a vibrating particle) is a timelike curve in space-time $M$. Indications of such a clock can mathematically be represented by a monotonically increasing function along this curve -- the time function for this clock. If there exists a single function $\T$ that is a time function for a family of clocks  filling the space-time $M$, such a function $\T$ is called a global time function. It measures a global time in the universe. As we can see, there exists a deep connection between global time and the above mentioned measurement stability property. It shows that doing physics automatically requires both: global time and stable causality.

We should notice that the spacelike hypersurfaces $\T$ = const give surfaces of simultaneity in the universe, but they are not unique. However, one can `synchronize the universe' by imposing a yet stronger causality condition. This is done in the following way. First, we define the Cauchy hypersurface of space-time $M$ as a hypersurface $S$ such that every inextendible timelike curve crosses $S$ exactly once. This definition was first introduced in the theory of partial differential equations \cite{Leray}. The initial data for a given hyperbolic PDE are given on a Cauchy surface. A space-time $M$ is globally hyperbolic if and only if it can be presented as a  product manifold  $M \simeq \sR \times S$, where $\sR $ is the range of a global time-function and $S$ a spacelike Cauchy surface in $M$ (see \cite{GerochSplitting,BS06}). 
The global time function $\T$, such that $\T^{-1}(t), \; t \in \T$ is a Cauchy surface, is called a Cauchy time function.

Globally hyperbolic space-time is the closest to the Newtonian absolute space we can get within the framework of general relativity. It is deterministic in the sense that initial data given on a Cauchy surface determine, in principle, the entire history of the universe. However, the so-called Cauchy problem in general relativity, in all its mathematical details, is far from being simple and easy (see \cite[Chapter 7]{HawkingEllis} or \cite{Ringstrom}).

\section{Time and Causality as Primitive Notions\label{sec:primitve}}

It is clear from the inspection of the axioms of general relativity that they presuppose some primitive notions of time and causality. In the EPS axiomatic system, echoes and messages are operationally defined in terms of coincidences of clock readings and acts of emission or reception of light rays. However, intuition smuggles into this operationistic picture an idea that the received echo is actually \emph{caused} by the radar signal. It seems, therefore, that some elementary notion of causality underlies all of the EPS system.

The same must be said about the primitive notion of time implicitly employed in the EPS axioms. Indeed, each history of a particle or a photon carries a $C^0$-structure which assures the local homeomorphism with $\sR $. This can be interpreted as a local flow of time with no preferred time orientation. On top of it, in order to make sense out of the `echoes and messages' axioms, one has to assume that the signal is emitted \emph{before} it is received. This means that each history of a particle or a photon is in fact a (totally) ordered collection of events with a distinguished future-direction. While the choice of a `future' and a `past' is purely conventional, fixing a consistent choice for all histories of particles amounts to assuming that the spacetime $M$ is time-oriented.

Therefore, the notion of local time flow and some elementary concept of causality are the minimum upon which a further hierarchy of stronger and stronger  conditions is built. As we have seen, the stable causality condition plays a special role in this hierarchy. It guarantees both the existence of global time and allows for stable measurement results.

All the structures described above, suitably synchronized with each other, are contained in the Lorentz metric structure. It is a standard result that the Lorentz metric structure  exists globally on a space-time manifold $M$ if and only if a non-vanishing direction field exists on $M$. Moreover, the Lorentz metric can always be chosen in such a way to make this direction field timelike \cite{Geroch71}. Since such a field on a space-time manifold locally always exists, the same is true for the Lorentz metric. The existence of a Lorentz metric is strictly related to the possibility of performing space and time measurements; therefore, it is almost synonymous with the possibility of doing physics. Above, we have identified such a possibility with the existence a local `topological time' (a $C^0$-structure on each history of a test particle); here we have the same condition raised to the metric level.\footnote{It is interesting to ask how this condition looks from the global point of view. The answer is that if $M$ is noncompact, such a nonvanishing direction field, and consequently a Lorentz metric, always exists, but if $M$ is compact it exists if and only if the Euler--Poincar\'e characteristic of $M$ vanishes \cite{GerochCompact}.}

A word of warning is needed at this point. Our conclusions are valid only within the conceptual framework of what we have called relativistic model of space time, and only within its reconstruction as it is presented above. Other axiomatic approaches to the geometry of space-time are possible (see, for, instance, \cite{Andreka,Covarrubias93,Guts95}) and they can give rise to different interpretations.\footnote{Different -- within certain limits. There is one important constraint: the mathematical structure of space-time must be preserved by all interpretations. One could say that the mathematical structure is `invariant' with respect to all admissible interpretations.} However, we should take into account the fact that it is the theory of general relativity that is deeply rooted in this model, and since this theory is very well founded on empirical data, it would be unwise to look for a different model (within the limits of its empirical verifications). We should also emphasise that the EPS axiomatic approach should not be easily replaced by other approaches since it renders justice, and does it very well, to both the `theoretical practice' of mathematical physicists and the operational demands of experimentalists (it has a strong (quasi)operationistic flavour).

The above analysis shows that temporal and causal properties are strongly coupled with each other, and one cannot say which is logically (or ontologically) prior with respect to the other. They are unified in the Weyl structure to provide a basis for the full dressed concept of time and causality. In this sense, causal theory of time \`a la Leibniz (causality implies time) is not supported by the relativistic model considered here. The same should be said about an attempt, \`a la Hume, to reduce causal interactions to merely a temporal succession. It should be taken into account that axiomatic systems can be composed in various ways: various concepts can be selected as primitive and various statements can be accepted as axioms of the system, depending on criteria one adopts. The EPS axiomatic system has an advantage over other axiomatic systems that it is quasi-operational, i.e. its axioms describe some simple empirical procedures, although they do so in a highly idealized way. We could conclude that, from the philosophical point of view, space-time is a rich holistic structure, and the choice of a specific axiomatics corresponds to the choice of the angle at which we contemplate the whole.

\section{\label{sec:prob}Probabilities in space-time}

The axiomatics of EPS is based on the assumption that photons and particles are point-like, because their histories $L$ and $P$ are collections of definite events. But the concept of localised particles is not supported by quantum theory \cite{Malament1996}. In contrast, quantum mechanics implies that `statistical predictions do apply to single events' \cite{QIandGR}. Actually, even within the classical theory the space-time location measurements are always affected by experimental uncertainties and, consequently, the events associated with them are somewhat `spread' in the space-time. How does this fact affect the notions of time and causation?

It is useful to distinguish two levels of conceptualisation in a physical theory: an `effective' one, which directly relates to the experimental data, and a `fundamental' one, which aims at providing an explanation for the data. At the effective level we deal with raw data, which correspond to elemental events, such as the click of a detector. The interpretation of these raw data requires a theoretical formalism, which associates them with physical phenomena. It is at this stage when the uncertainties, inflicted by experimental limitations or, possibly, by fundamental randomness inherent in the adopted theory, arise. In consequence, while elemental experimental events are definite, the final outcome acquires a probabilistic form, which involves both systematic and statistical errors of the measured quantities. It applies, in particular, to  space-time location measurements.

The concepts of time and causality stemming from the EPS axiomatisation can only be applied at the effective level with definite events. In order to extend them to the `fundamental' level one needs to take into account the probabilistic nature of experimental outcomes.

A natural universal mathematical structure suitable to grasp the uncertainty of events is provided by the probability measures. Given a space-time $M$ one defines the space $\P(M)$ of all probability measures\footnote{Technically, one should assume that the measures in $\P(M)$ are Borel, to assure compatibility with the topology of $M$. Furthermore, since any relativistic space-time is a Polish space, all elements of $\P(M)$ are actually Radon --- see \cite{AHP2017} for the details.} on $M$. An element $\mu \in \P(M)$ is a function, which associates with every (measurable) region of space-time $M$ a probability, i.e. a number from the interval $[0,1]$. Given any such region $K$, the number $\mu(K)$ specifies the probability of the occurrence of \emph{some} event associated with $K$ and $\mu$. For instance, if $\mu$ models the response of a detector, which occupies a volume $V$ in space and operates within a time-interval $T$, then $\mu(T \times V)$ gives the probability of a single detector's click. More precisely, the number $\mu(T \times V)$ answers the operational question: what is the probability of the signal detection \emph{if} a suitable detector operating in space-time region $T \times V$ is placed. If the detector does click, then we interpret it as registration of the signal to which it was tuned, coming with a space-time label determined by the region $K = T \times V$. If the detector does not click, then it provides us with a definite information that the signal is with certainty \emph{somewhere outside} of the region $K$. The latter should be distinguished from the situation in which the detector is not switched on, as the `unperformed experiments have no results' \cite{NoResults}.

Clearly, any physical device has some space- and time-resolution, so that the region $K \subset M$ cannot consist of a single point. On the other hand, the measure itself 
can in principle be localised acutely. Indeed, given any point $p \in M$ one can associate with it a Dirac measure $\delta_p \in \P(M)$. The latter would yield 1 whenever the event $p$ lies in the space-time region covered by the detector, $p \in K$, and 0 if this is not the case. The Dirac measures can thus serve to model the classical --- localised --- particles. Note, however, that even in this case the finite size of the detector induces an uncertainty of ascribing a definite space-time label, as the click only gives the information that the event $p$ happened \emph{somewhere} within the space-time region $K$.

We see that the probabilistic nature of experiments coerces a reinterpretation of the relativistic space-time $M$ itself. The points in $M$ should not be seen as actual events, but rather $M$ serves merely as a support-space for the probability measures. The actual events arise at an effective level, as a result of an interaction between a localised detector and a `particle'. Such an understanding of the measuring process is usually put forward in the context of quantum theory (cf. the discussion in \cite[Section I.1]{Haag} or \cite{QIandGR}), but in fact it is a general operational concept, not relying on how we eventually model the `particles' (as well as `detectors' and `interactions') --- cf. \cite{PRA2020}. Consequently, there are no events on the fundamental level in a probabilistic physical theory.

\section{\label{sec:prob_caus}Causality of probabilities}

\begin{figure}[h]
\begin{center}
\includegraphics[scale=0.2]{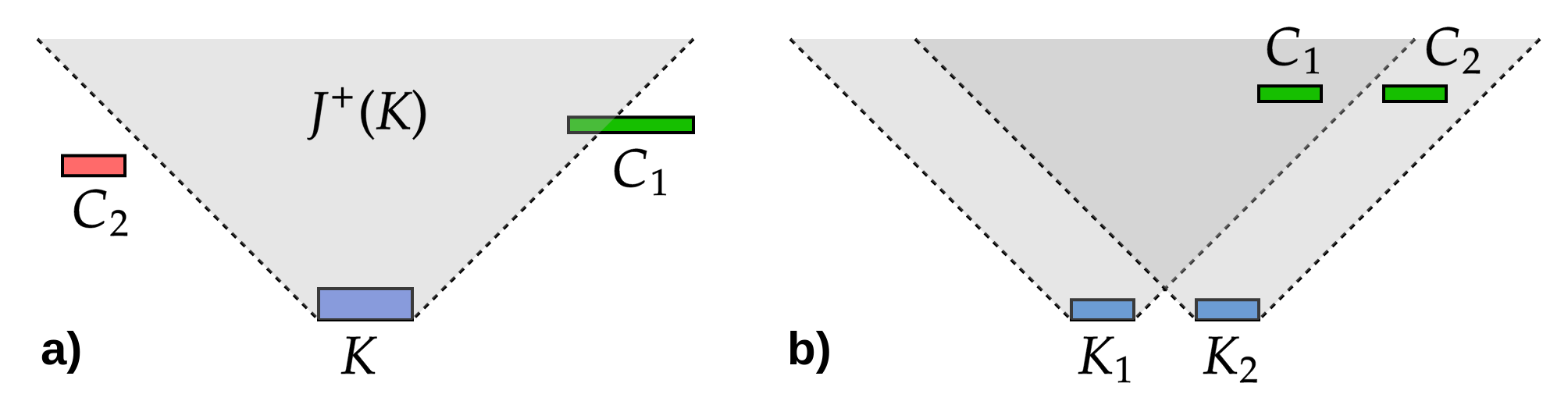}
\caption{\label{Fig2} \textbf{a)} The device localised in region $K$ can causally influence the device in $C_1$, but not in $C_2$. \textbf{b)} Two pairs of detectors, localised in regions $K_1$, $K_2$ and $C_1$, $C_2$, respectively, respond to the signal modelled by measures $\mu$ and $\nu$, respectively. With the corresponding probabilities of detection, $\mu(K_1) = a$, $ \mu(K_2) = 1-a$, $\nu(C_1) = b$ and $\nu(C_2) = 1-b$, we have $\mu \preceq \nu$ if and only if $a \leq b$. Intuitively, if $a$ would be greater than $b$, then the probability would have to `leak out' of the causal future $J^+(K_1)$.}
\end{center}
\end{figure}

The relativistic causal structure discussed in Section \ref{sec:global} pertained to definite point-like events. It is fairly straightforward to extend it to a more realistic situation when signal-processing devices are characterised by an extended region $K=T \times V$ within the space-time. Indeed, one can simply say that the device operating in $K$ can causally affect the device in $C$ if and only if $J^{+}(K) \cap C \neq \emptyset$. In other words, if the second device is localised outside of the causal future of the first device, then there can be no causal influence of the first one on the second (see Fig. \ref{Fig2} a)).

Such an idea employs only signal-processing devices and does not refer to the nature of the signals themselves. Moreover, it does not take into account the probabilistic nature of the devices' outcomes. Suppose that we have two pairs of localised detectors as in Fig. \ref{Fig2} b). The detectors localised in regions $K_1$, $K_2$, $C_1$ and $C_2$ click with probabilities $\mu(K_1) = a$, $ \mu(K_2) = 1-a$, $\nu(C_1) = b$ and $\nu(C_2) = 1-b$, respectively. When can the detection statistics $\mu$ causally influence the detection statistics $\nu$?

A rigorous answer to this question is provided within the formalism established recently by Eckstein and Miller \cite{AHP2017}. They proposed the following definition \footnote{Strictly speaking, this definition was presented in another work of Eckstein and Miller \cite{PRA2017}, as a refinement of one of the equivalent characterisation of casual precedence for probability measures established in \cite[Theorem 8]{AHP2017}. The same definition was independently put forward by Stefan Suhr in \cite{Suhr2016}.}: For two probability measures $\mu, \nu \in \P(M)$ on a given space-time $M$ we say that $\mu$ causally precedes $\nu$, symbolically $\mu \preceq \nu$, if for every compact\footnote{Working solely with compact subsets of $M$ is not a limitation, because $\mu(X)$ is determined for any measurable $X$ through the formula $\mu(X) = \sup \{ \mu(K) \, \vert \, K \subset X, \, K \text{ compact} \}$, which is valid for Radon measures. Also, on the physical side, one can safely assume that the space-time region associated with any device is compact, i.e. closed and bounded.} set $K \subset \supp \mu$ we have $\mu(K) \leq \nu (J^+(K))$. The general intuition behind this definition is that, for any compact region of space-time $K \subset M$ the probability cannot `leak out' of the causal future of $K$. If we apply this definition to the example depicted in Fig. \ref{Fig2} b), we conclude that the first pair of detectors can causally influence the second pair only if the relevant probabilities satisfy $a \leq b$.

Such a notion of causality for probabilities might seem purely formal, but it is in fact coherent with the operational `no-signalling principle' (see \cite{PR_box,Bell_nonlocal} and \cite{PRA2020}). The latter says that no actions or events in a space-time region $K$ can causally influence any measurement statistics outside of the causal future $J^+(K)$. For if they did, the information could be transferred superluminally by orchestrating a protocol, in which an observer in $K$ prepares multiple signal carriers and sends them to another observer outside of $J^+(K)$, who can read-out the information from his detection statistics. Such a protocol, when executed by pair of mutually travelling inertial observers, leads to causal loops  and consequent logical paradoxes --- see \cite{PRA2020} for a detailed discussion.

It turns out that the causal precedence between probability measures admits another equivalent \footnote{The equivalence of these two definitions holds, \cite[Theorem 8]{AHP2017}, in causally simple space-times, which have slightly weaker causal properties than the globally hyperbolic ones --- see e.g. \cite{MS08}.} characterisation: A measure $\mu$ causally precedes $\nu$ if and only if there exists a joint measure $\omega \in \P(M \times M)$ such that $\omega( \cdot \times M) = \mu( \cdot)$, $\omega( M \times \cdot) = \nu( \cdot)$ and $\omega (J^+) = 1$, where the set $J^+$ consists of all pairs of points $(p,q) \in M \times M$ such that $p \preceq q$. This condition is inspired by the optimal transport theory. It encodes the following intuition (cf. \cite{PRA2017}): `Each infinitesimal part of a probability distribution must travel along a future-directed casual curve.'

\section{\label{sec:prob_evo}Time-evolution of probabilities}

In the previous section we have seen how the causal order of points in a relativistic space-time can be extended to probability measures modelling detection statistics. What about the time-evolution?

Firstly, let us recall that within the relativistic setting there is no preferred time parameter and one needs to treat the space-time as a global holistic structure. Time-evolution arises when an observer adopts a local coordinate chart, for instance by adopting the signals-and-echoes method of EPS. On the other hand, as discussed in Section \ref{sec:global}, space-times with sufficiently robust causal structures admit \emph{global} time functions, $\T: M \to \R$, which determine the hypersurfaces of simultaneity. However, the choice of a time function is by no means unique. Indeed, even the simplest Minkowski space-time admits a continuous family of equivalent time-slicings associated with different inertial observers. More generally, the choice of a time-function can be associated with a `global observer', i.e. a collection of observers, parameterised by points on an achronal hypersurface, each of which follows some future-directed time-like curve. In the case of an inertial slicing of the Minkowski space-time, all such observers travel in parallel --- in the same direction and with the same speed. In general, this need not be the case and the collection of all admissible time-slicings is vast.

The measures on space-time are inherently non-local objects. They can, in principle, be spread throughout the entire space-time. The latter situation arises, for instance, if one associates measures with the quantum states emerging from quantum field theory \cite{ReehSchlieder,Malament}. For this reason, if one wants to consider the time-evolution of general measures one needs to adopt the global perspective of time-slicings, rather than the local one based on coordinate charts.

Let us then fix a time-function $\T: M \to \R$ on a given space-time $M$. In order to guarantee the well-posedness of the time-evolution problems, we shall assume that the space-time $M$ is globally hyperbolic and that the level-sets of $\T$ are Cauchy hypersurfaces, which means that they can accommodate initial data for some hyperbolic evolution equation (see e.g. \cite{Ringstrom}). Let us emphasise here that this is a limitation only in the cosmic context when we interpret the space-time $M$ as the whole Universe. Locally, any space-time, even one containing closed time-like curves, is always globally hyperbolic in some open neighbourhood of any point \cite{Penrose1972}. One can thus safely consider models of  phenomena, which are non-local up to some scales. The truly global ones, for instance impelled by quantum field theory, require the (rather standard, see e.g. \cite{curvedQFT_rev}) assumption of global hyperbolicity of the entire Universe.

Having chosen a time-function $\T$ we obtain the corresponding splitting $M \simeq \R \times S$ with a Cauchy hypersurface $S$. Consider now a family of probability measures supported on subsequent time-slices, $\mu_t \in \P(S_t)$, with $S_t \vc \{t\} \times S$. Such objects are thus `localised in time', but `delocalised in space'. The time-evolution of measures, for a chosen global time-function $\T$, is the family $\{\mu_t\}_{t}$. Equivalently, one can define it as a map $t \mapsto \mu_t \in \P(M)$, with $\supp \mu_t \subset S_t$.

For a given time function $\T$ the quantity $\{\mu_t\}_t$ models a time-evolution of probabilities \emph{potentially} registered by an observer associated with $\T$. Note that the uncertainty of the time-moment of the detection comes solely from the finite time-resolution of the measuring device, characterised by an interval $T$. The measures themselves are `localised in time'. This means that, within this framework, there are no \emph{a priori} limits on the time-resolution of the measurements. Such an assumption is met both in general relativity and in quantum theory, though it might fail in some quantum gravity theories \cite{Minimal_Length_Review}.

Observe that the notion of evolution of measures includes that of a trajectory of point-like particles. Indeed, with any parametrised (piece-wise smooth) curve $t \mapsto \gamma(t) \in \M$, for a chosen time-parameter, one can associate the measures $\mu_t = \delta_{\gamma(t)}$, which are localised both in time and in space. Recall that such a trajectory of a point-like particle is (future-directed) causal if $\gamma(s) \preceq \gamma(t)$ for all $s \leq t$. In the same vein, one says that a general evolution of measures $\{\mu_t\}_t$ is causal if $\mu_s \preceq \mu_t$ for all $s \leq t$. 

The trajectory of a point-like particle is in fact an observer-independent quantity. Indeed, every such trajectory can be deparametrised, i.e. treated as a worldline --- the collection $[\gamma]$ of points in $M$. This fact facilitates a sound interpretation of trajectories: different observers can use different parametrisations, determined by their local frames, of \emph{the same} moving particle. In other words, the travelling point-like particle itself is an observer-independent entity, while the description of its motion requires some time-parametrisation. The existence of an invariant object --- the worldline --- guarantees that different parameterisations of the trajectory are \emph{covariant}. That is, there exists a unequivocal prescription allowing two different observers, using different time-parametrisation, to compare the outcomes of their observation of the same moving point-like particle.  

As it turns out \cite{Miller17a}, for a general causal evolution of measures on a globally hyperbolic space-time $M$ there also exists an invariant object, i.e. an object independent of the choice of the time function. In order to unveil it one needs, firstly,  to consider the space of all future-directed causal curves on $C^I_\T(M)$, parametrised in accordance with the global time-function $\T$, for some time-interval $I \subseteq \R$ and endow it with a suitable (compact-open) topology, which turns it into a locally compact Polish space\footnote{Polish spaces are separable and completely metrisable, what already makes them suitable for developing much of the abstract probability theory.}. Then, one proves \cite[Theorem 1]{Miller17a} that an evolution of measures $\{\mu_t\}_t$ is causal if and only if there exists a measure $\sigma \in \P(C^I_\T)$ such that $(\ev_t)_{\#} \sigma = \mu_t$ for any $t \in I$, where $\ev$ is the standard evaluation map: $\ev_t (\gamma) = \gamma(t) \in M$ for any curve $\gamma \in C^I_\T(M)$. Secondly, one shows that if $I = \R$ then every such $\sigma$ is in one-to-one correspondence with a measure $[\sigma] \in \P(\mathcal{C}(M))$ on the space of worldlines $\mathcal{C}(M)$ in $M$. In this way, one arrives at an object $[\sigma]$ --- a probability measure on worldlines --- which is manifestly invariant.

If the evolution of measures describes the motion of a single point-like particle, $\mu_t = \delta_{\gamma(t)}$, then the associated measure $[\sigma]$ is supported on a single worldline $[\gamma]$, ie. $[\sigma] = \delta_{[\gamma]}$. However, for a general evolution of measures the object $\sigma$, and hence $[\sigma]$, is \emph{not unique}. This can be seen from the optimal-transport perspective on the evolution of measures. Indeed, for any two fixed probability distributions on time-slices $S_s$ and $S_t$ there might exist many causal paths, which transport the infinitesimal portions of probablity from $S_s$ to $S_t$. Nevertheless, thanks to the existence of an invariant object, any causal time-evolution of measures is \emph{covariant}, in the sense that one has a precise prescription on how to translate an evolution of measures as witnessed form the perspective of one global time-function $\T_1$ to the perspective of another $\T_2$. In particular, any two observers will agree upon the (a)causality of an evolution of measures. For a detailed discussion see \cite{Miller17a} and also\footnote{Miller has recently shown that there exists yet another invariant object associated with every causal evolution of measures, which is related to the continuity equation.} \cite{Miller21}.

The concept of the causal evolution of probability measures is useful to inspect the dynamical equations, in classical, quantum or `post-quantum' theories, for their compatibility with the causal structure of a relativistic space-time \cite{PRA2020}. In particular, the (normalised) electromagnetic energy density $u \vc \tfrac{1}{2 \mathcal{E}} \big( \varepsilon_0 \norm{\mathbf{E}}^2 + \tfrac{1}{\mu_0} \norm{\mathbf{B}}^2 \big)$, with a finite total energy $\mathcal{E} = \int_{\mathbb{R}^3} u(0,x) dx$,  yields a family of probability measures $\mu_t \vc u(t,x) \, d x$ on $\mathbb{R}^3$, which evolves causally. The same is true for the probability density $\psi^{\dagger}(t,x) \psi(t,x) \, d x$ resulting from a wave-function $\psi(t,x)$ evolving according to the Dirac equation, possibly including external gauge potentials \cite{PRA2017}. On the other hand, the evolution of a probability density associated with a quantum wave function driven by the Schr\"odinger equation with a positive definite Hamiltonian is typically \emph{not} causal \cite{PRA2017} (see also \cite{Hegerfeldt1,Hegerfeldt1985}). Such an incompatibility with the relativistic causality can be quantified in terms of the characteristic time- and length-scales \cite{PRA2017} and can be utilised to restrain certain `post-quantum' theories \cite{PRA2020}.

The philosophical picture that emerges from these considerations is the following: With every physical phenomenon one can associate a mathematical object --- a probability measure on the space of worldlines in a given space-time $M$. This object is \emph{global}, which means that it provides a holistic model for the entire history of a given physical system and thus we have no direct access to it. The time-evolution of this system is an emergent concept, which arises when an observer chooses a global time-function $\T$ and studies its local properties by registering the time-evolution of statistics $\mu_t(T \times V)$ with the help of a device localised in the volume $V$ and having a time-resolution $T$. A different observer using a different time-function would register \emph{different} statistics $\mu_t'(T' \times V')$, when measuring \emph{the same} physical system with \emph{the same} detecting device. Yet, the two observers can consistently compare their results by translating the registered probabilities through the invariant object.

\section{\label{sec:Conclusions}Conclusions}

Our analysis shows that some primitive notions of `time', `space' and `causality' are coerced by the very methodology of science. Indeed, in any experimental setup one needs to specify an `input', which always comes \emph{before} the `output'. In the same vein, some primitive notion of space is presupposed by the very fact that any experimental setup is a physical device, which occupies some definite volume and needs to be placed and oriented --- e.g. a telescope pointed towards a specific star.

More formally, an experiment is in essence a collection of data $\{d_i\}$. (Without any loss of generality we can thus assume that $d_i$'s are just bits.) Any such bit is \emph{local}, that is uniquely associated with an actual event pertaining to the physical world. Regardless of the adopted theoretical framework, we are bound to place the bits in `space' and `time'. The very notation $b_1, b_2$ informs one that the bit $b_1$ has been input, acquired or communicated before the bit $b_2$. If, for any reason, we are dealing with different data sets, we distinguish them by using different labels, say $a$ and $b$. This leads to a primitive notion of space: The event $a_1$ could have taken place simultaneously with $b_1$, but at a different `place'. 

Furthermore, we always make some presuppositions about the causal relations among the data. In particular, we always assume that the input data influenced the output, but not the other way round (cf. e.g. Axiom 1 `causality' in \cite{Chiribella2011}). The same must be said about the basic probabilistic structures. Indeed, any viable analysis of experimental data must be endowed with probabilistic concepts such as relative frequencies of the outcomes and correlations among the data sets. It is clear that the very methodology of physics requires the primitive probabilistic concepts, on top of the space-time-related ones.

Every physical theory is based on some mathematical structures, which \emph{per se} are purely formal objects. In order to relate theories with actual experiments, one needs a `minimal' operational interpretation. The latter bridges between the primitive notions of space, time, causation and the rigorous mathematical notions modelling a chosen phenomenon. Moreover, any physical theory fosters probabilistic predictions, which can be verified against empirical data. This means, in particular, that a theory must provide a mechanism explaining the relative frequencies and correlations observed in a relevant experiment, in terms of the modelled physical system.

While the testing of a physical model necessarily requires the use of some primitive notions, a valid physical theory may refine some of these notions and unveil some unexpected connections between them. In this vein, general relativity has revealed, via the Hawking theorem, a deep connection between the existence of a global cosmic time and stability of space-time measurements. On the other hand, quantum theory implies that the measurement outcomes are afflicted by ontic uncertainties, not related to our subjective lack of knowledge about the studied system. 

These conclusions clearly hinge upon the adopted primitive notions in the first place, which are necessary for the interpretation of empirical data. Nevertheless, the `unreasonable effectiveness' of the mathematical-empirical methodology \cite{Wigner} in modelling natural phenomena strongly suggests that a valid physical theory reveals some ontological aspects, which underlie the primitive concepts of time, causality and probability. It also suggests that the world should be contemplated from a holistic perspective, one in which time, causality and probability are irreducibly related. This calls for the development of a new philosophical discourse, going beyond the classical dichotomy \`a la Leibniz versus Hume, which would be able grasp the integral connections between the pillars of the methodology of physics. 
	
\section*{Acknowledgements}

We are grateful to Tomasz Miller for his thoughtful comments on the manuscript. M.E. would like to thank Pawe{\l} Horodecki for the numerous inspiring discussions on causality and randomness.

\end{document}